\documentclass[letterpaper, 10 pt, conference]{ieeeconf}
\IEEEoverridecommandlockouts

\usepackage{amsfonts,amsmath,amssymb} 

\def\rbb{\mathbb{R}}
\def\trp{^T}

\def\half{\frac{1}{2}}
\usepackage{psfrag,color}
\usepackage{enumerate,cite,latexsym,graphicx}

\title{\LARGE \bf Implementation of a Direct Coupling Coherent Quantum Observer including Observer Measurements}

\author{Ian R.~Petersen and Elanor H. Huntington
\thanks{This work was supported by the
Australian Research Council (ARC) and the Chinese Academy of Sciences President’s International Fellowship Initiative (No. 2015DT006). }%
\thanks{Ian R. Petersen is with the School of  Engineering and Information Technology, 
        University of New South Wales at the Australian Defence Force Academy, Canberra ACT 2600, Australia.
         {\tt\small i.r.petersen@gmail.com} } 
\thanks{Elanor H. Huntington is with the 
College of Engineering and Computer Science, The Australian National University, Canberra, ACT 0200,
Australia. Email: Elanor.Huntington@anu.edu.au.}
}%

\begin{document}

\maketitle
\thispagestyle{empty}
\pagestyle{empty}

\begin{abstract}
This paper considers the problem of constructing a direct coupling quantum observer for a quantum harmonic oscillator system. The proposed observer is shown to be able to estimate one but not both of the plant variables and produces a measureable output using homodyne detection. 
\end{abstract}

\section{Introduction} \label{sec:intro}
A number of papers have recently considered the problem of constructing a coherent quantum observer for a quantum system; e.g., see \cite{MJ12a,VP9a,EMPUJ6a,PET14Aa}. In the coherent quantum observer problem, a quantum plant is coupled to a quantum observer which is also a quantum system. The quantum observer is constructed to be a physically realizable quantum system  so that the system variables of the quantum observer converge in some suitable sense to the system variables of the quantum plant. The papers \cite{PET14Aa,PET14Ba,PET14Ca,PET14Da}  considered the problem of constructing a direct coupling quantum observer for a given closed quantum system. In \cite{PET14Aa}, the proposed observer is shown to be able to estimate some but not all of the plant variables in a time averaged sense. Also, the paper \cite{PeHun1a} shows that a possible experimental implementation of the augmented quantum plant and quantum observer system considered in \cite{PET14Aa} may be constructed using a non-degenerate parametric amplifier (NDPA) which is coupled to a beamsplitter by suitable choice of the NDPA and beamsplitter parameters. 

One important limitation of the direct coupled quantum observer results given in \cite{PET14Aa,PET14Ba,PET14Ca,PET14Da,PeHun1a} is that both the quantum plant and the quantum observer are closed quantum systems. This means that it not possible to make an experimental measurement to verify the properties of the quantum observer. In this paper, we address this difficulty by extending the results of \cite{PET14Aa} to allow for the case in which the quantum observer is an open quantum linear system whose output can be monitored using homodyne detection. In this case, it is shown that similar results can be obtained as in \cite{PET14Aa} except that now the observer output is subject to a white noise perturbation. However, by suitably designing the observer, it is shown that the level of this noise perturbation can be made arbitrarily small (at the expense of slow observer convergence). Also, the results of \cite{PeHun1a} are extended to show that a possible experimental implementation of the augmented quantum plant and quantum observer system  may be constructed using a non-degenerate parametric amplifier (NDPA) which is coupled to a beamsplitter by suitable choice of the NDPA and beamsplitter parameters. In this case, the NDPA contains an extra field channel as compared to the result in \cite{PeHun1a} and this extra channel is used for homodyne detection in the observer.

\section{Direct Coupling Coherent Quantum Observer with Observer Measurement}
\label{sec:observer}
In this section, we extend the theory of \cite{PET14Aa} to the case of a direct coupled quantum observer which is also coupled to a field to enable measurements to be made on the observer. In our proposed direct coupled coherent quantum observer, the quantum plant is a single quantum harmonic oscillator which is a linear quantum system (e.g., see \cite{JNP1,PET10Ba,NJD09,GJ09,ZJ11}) described by the non-commutative differential equation
\begin{eqnarray}
\dot x_p(t) 
&=& 0; \quad x_p(0)=x_{0p}; \nonumber \\
z_p(t) &=& C_px_p(t)
 \label{plant}
\end{eqnarray}
where $z_p(t)$ denotes the  system variable to be estimated by the observer and $C_p\in \rbb^{1 \times 2}$. 
 This quantum plant corresponds to a plant Hamiltonian
$\mathcal{H}_p=0$. Here $x_p = \left[\begin{array}{l}q_p\\p_p\end{array}\right]$ where
$q_p$ is the plant position operator and $p_p$ is the plant momentum operator. It follows from (\ref{plant}) that the plant system variables $x_p(t)$ will remain fixed if the plant is not coupled to the observer.

\begin{figure}[htbp]
\begin{center}
\includegraphics[width=8cm]{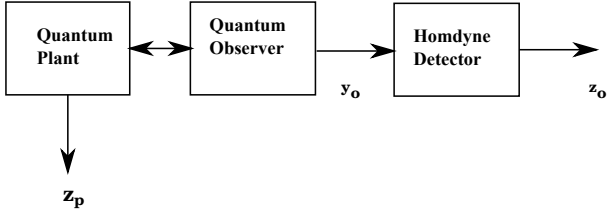}
\end{center}
\caption{Plant Observer System.}
\label{F1}
\end{figure}

We now describe the linear quantum system  which will correspond to the quantum observer; see also \cite{JNP1,PET10Ba,NJD09,GJ09,ZJ11}. 
This system is described by a quantum stochastic differential equation (QSDE) of the form
\begin{eqnarray}
dx_o &=& A_ox_odt+B_odw;\quad x_o(0)=x_{0o};\nonumber \\
dy_o &=& C_ox_odt+dw;\nonumber \\
z_o(t) &=& Ky_o
 \label{observer}
\end{eqnarray}
where $dw = \left[\begin{array}{l}dQ\\dP\end{array}\right]$ is a $2\times 1$ vector of quantum noises expressed in quadrature form corresponding to the input field for the observer and $dy_o$ is the corresponding output field; e.g., see \cite{JNP1,NJD09}.  The observer output $z_o(t)$ will be  a real scalar quantity obtained by applying homodyne detection to the observer output field. $A_o \in \rbb^{2\times 2}$, $B_o \in \rbb^{2\times 2}$, $C_o\in \rbb^{2 \times 2}$.  Also,  $x_o(t)=\left[\begin{array}{l}q_o\\p_o\end{array}\right]$  is a vector of self-adjoint 
system variables corresponding to the observer position and momentum operators; e.g., see \cite{JNP1}. We  assume that the plant variables commute with the observer variables. The system dynamics (\ref{observer}) are determined by the observer system Hamiltonian and coupling operators which are operators on the underlying  Hilbert space for the observer. For the quantum observer under consideration, this Hamiltonian is a self-adjoint operator given by the 
quadratic form:
$\mathcal{H}_o=\half x_o(0)\trp R_o x_o(0)$, where $R_o$ is a real symmetric matrix. Also, the coupling operator $L$ is defined by a matrix $W_o \in \rbb^{2\times 2}$ so that
\begin{equation}
\label{Wo}
\left[\begin{array}{c} L+ L^* \\ \frac{ L - L^*}{i}\end{array}\right]
 = W_o x_o.
\end{equation}
Then, the corresponding matrices $A_o$, $B_o$ and $C_o$ in 
(\ref{observer}) are given by 
\begin{equation}
\label{eq_coef_cond_Ao}
A_o=2J R_o+\frac{1}{2}W_o^TJW_o, ~~B_o = JW_o^TJ,~~C_o = W_o
\end{equation}
 where 
\[
J = \left[\begin{array}{ll} 0 & 1  \\ -1 & 0 \end{array} \right];
\]
e.g., see \cite{JNP1,NJD09}.  
 Furthermore, we will assume that the  quantum observer  is coupled to the quantum plant as shown in Figure \ref{F1}. 
In addition, we define a coupling Hamiltonian which defines the coupling between the quantum plant and the  quantum observer:
\[
\mathcal{H}_c = x_{p}(0)\trp R_{c} x_{o}(0).
\]
The augmented quantum linear system consisting of the quantum plant and the quantum observer is then a linear quantum system  described by the total Hamiltonian
\begin{eqnarray}
\mathcal{H}_a &=& \mathcal{H}_p+\mathcal{H}_c+\mathcal{H}_o\nonumber \\
 &=& \half x_a(0)\trp R_a x_a(0)
\label{total_hamiltonian}
\end{eqnarray}
where
\begin{equation}
\label{xaRa}
x_a = \left[\begin{array}{l}x_p\\x_o\end{array}\right],~~
R_a = \left[\begin{array}{ll}0 & R_{c} \\
R_{c}^T & R_{o}
\end{array}\right],
\end{equation}
and the coupling operator $L$ defined in (\ref{Wo}). Extending the approach used in \cite{PET14Aa}, we assume that we can write
\begin{equation}
\label{Rc}
R_c = \alpha\beta^T,
\end{equation}
 $R_o = \omega_o I$, $W_o = \sqrt{\kappa} I$ where $\alpha  \in \rbb^{2}$, $\beta \in \rbb^{2}$,  $\omega_o > 0$ and $\kappa > 0$. In addition, we assume
\begin{equation}
\label{alpha}
\alpha =  C_p^T.
\end{equation}
Then, we can write the QSDEs describing the closed loop system as follows:
\begin{eqnarray}
\label{augmented1}
d x_p &=& 2 J \alpha\beta^T x_o dt;\nonumber \\
d x_o &=& 2\omega_o J x_o dt + 2J\beta\alpha^T x_p dt + \frac{1}{2} JW_o^TJW_o x_o dt\nonumber \\
&&+ JW_o^TJdw \nonumber \\
&=& \left[\begin{array}{ll} -\frac{\kappa}{2} & 2 \omega_o \\-2\omega_o &  -\frac{\kappa}{2} \end{array}\right]x_o dt 
+2J\beta \alpha^T x_p dt - \sqrt{\kappa}dw; \nonumber \\
dy_o &=& W_ox_o dt + dw \nonumber \\
&=& \sqrt{\kappa} x_o dt + dw;
\end{eqnarray}
e.g., see \cite{JNP1,NJD09}. Now it follow from (\ref{plant}) and (\ref{alpha}) that
\[
z_p = \alpha^T x_p. 
\]
Hence, it follows from the first equation in (\ref{augmented1}) that
\[
dz_p = 2 \alpha^T J \alpha\beta^T x_o dt = 0.
\]
That is, the quantity $z_p$ remains constant even after the quantum plant is coupled to the quantum observer. In addition, we can re-write the remaining equations in (\ref{augmented1}) as
\begin{eqnarray}
\label{augmented2}
d x_o &=& \left[\begin{array}{ll} -\frac{\kappa}{2} & 2 \omega_o \\-2\omega_o &  -\frac{\kappa}{2} \end{array}\right]x_o dt 
+2J\beta z_p dt - \sqrt{\kappa}dw; \nonumber \\
dy_o &=& \sqrt{\kappa} x_o dt + dw;
\end{eqnarray}

To analyse the system (\ref{augmented2}), we first calculate the steady state value of the quantum expectation of the observer variables as follows:
\begin{eqnarray*}
<\bar x_o> &=& -2 \left[\begin{array}{ll} -\frac{\kappa}{2} & 2 \omega_o \\-2\omega_o &  -\frac{\kappa}{2} \end{array}\right]^{-1} J\beta z_p \nonumber \\
&=& \frac{4}{\kappa^2+16 \omega_o^2}\left[\begin{array}{ll} \kappa & 4 \omega_o \\-4\omega_o &  \kappa \end{array}\right]J\beta z_p.
\end{eqnarray*}
Then, we define the quantity
\[
\tilde x_o = x_o - <\bar x_o> = x_o - \frac{4}{\kappa^2+16 \omega_o^2}\left[\begin{array}{ll} \kappa & 4 \omega_o \\-4\omega_o &  \kappa \end{array}\right]J\beta z_p.
\]
We can now re-write the equations (\ref{augmented2}) in terms of $\tilde x_0$ as follows
\begin{eqnarray}
\label{augmented3}
d \tilde x_o &=& \left[\begin{array}{ll} -\frac{\kappa}{2} & 2 \omega_o \\-2\omega_o &  -\frac{\kappa}{2} \end{array}\right]x_o dt 
+2J\beta z_p dt - \sqrt{\kappa}dw \nonumber \\
&=&  \left[\begin{array}{ll} -\frac{\kappa}{2} & 2 \omega_o \\-2\omega_o &  -\frac{\kappa}{2} \end{array}\right]\tilde x_o dt \nonumber \\
&&- 2 \left[\begin{array}{ll} -\frac{\kappa}{2} & 2 \omega_o \\-2\omega_o &  -\frac{\kappa}{2} \end{array}\right]
\left[\begin{array}{ll} -\frac{\kappa}{2} & 2 \omega_o \\-2\omega_o &  -\frac{\kappa}{2} \end{array}\right]^{-1} J\beta z_p dt\nonumber \\
&&+2J\beta z_p dt - \sqrt{\kappa}dw \nonumber \\
&=& \left[\begin{array}{ll} -\frac{\kappa}{2} & 2 \omega_o \\-2\omega_o &  -\frac{\kappa}{2} \end{array}\right]\tilde x_o dt 
 - \sqrt{\kappa}dw; \nonumber \\
dy_o &=& \sqrt{\kappa} \tilde x_o dt - 2 \sqrt{\kappa}\left[\begin{array}{ll} -\frac{\kappa}{2} & 2 \omega_o \\-2\omega_o &  -\frac{\kappa}{2} \end{array}\right]^{-1} J\beta z_p dt+ dw \nonumber \\
&=& - 2 \sqrt{\kappa}\left[\begin{array}{ll} -\frac{\kappa}{2} & 2 \omega_o \\-2\omega_o &  -\frac{\kappa}{2} \end{array}\right]^{-1} J\beta z_p dt + dw^{out}
\end{eqnarray}
where
\[
dw^{out} = \sqrt{\kappa} \tilde x_o dt + dw.
\]

We now look at the transfer function of the system 
\begin{eqnarray}
\label{augmented4}
d \tilde x_o &=& \left[\begin{array}{ll} -\frac{\kappa}{2} & 2 \omega_o \\-2\omega_o &  -\frac{\kappa}{2} \end{array}\right]\tilde x_o dt 
 - \sqrt{\kappa}dw; \nonumber \\
dw^{out} &=& \sqrt{\kappa} \tilde x_o dt + dw
\end{eqnarray}
which is given by
\[
G(s) = -\kappa \left[\begin{array}{ll}s+\frac{\kappa}{2} & - 2 \omega_o\\2\omega_o & s+\frac{\kappa}{2}\end{array}\right]^{-1}.
\]
It is straightforward to verify that this transfer function is such that
\[
G(j\omega)G(j\omega)^\dagger =I
\]
for all $\omega$. That is $G(s)$ is all pass. Also, the matrix $\left[\begin{array}{ll} -\frac{\kappa}{2} & 2 \omega_o \\-2\omega_o &  -\frac{\kappa}{2} \end{array}\right]$ is Hurwitz and hence, the system (\ref{augmented4}) will converge to a steady state in which $dw^{out}$ represents a standard quantum white noise with zero mean and  unit intensity. Hence, at steady state, the equation 
\begin{equation}
\label{yo}
dy_o = - 2 \sqrt{\kappa}\left[\begin{array}{ll} -\frac{\kappa}{2} & 2 \omega_o \\-2\omega_o &  -\frac{\kappa}{2} \end{array}\right]^{-1} J\beta z_p dt + dw^{out}
\end{equation}
shows that the output field converges to a constant value plus zero mean white quantum noise with unit intensity. 

We now consider the construction of the vector $K$ defining the observer output $z_o$. This vector determines the quadrature of the output field which is measured by the homodyne detector. We first re-write equation (\ref{yo}) as
\[
dy_o =  ez_p dt + dw^{out}
\]
where 
\begin{equation}
\label{e}
e = - 2 \sqrt{\kappa}\left[\begin{array}{ll} -\frac{\kappa}{2} & 2 \omega_o \\-2\omega_o &  -\frac{\kappa}{2} \end{array}\right]^{-1} J\beta
\end{equation}
is a vector in $\rbb^{2}$. Then
\[
dz_o = Kez_pdt + Kdw^{out}.
\]
Hence, we choose $K$ such that
\begin{equation}
\label{Kconstraint}
Ke = 1
\end{equation}
and therefore 
\[
dz_o = z_pdt + dn
\]
where 
\[
dn = K dw^{out}
\]
will be a white noise process at steady state with intensity $\|K\|^2$. Thus, to maximize the signal to noise ratio for our measurement, we wish to choose $K$ to minimize $\|K\|^2$ subject to the constraint (\ref{Kconstraint}). Note that it follows from (\ref{Kconstraint}) and the Cauchy-Schwartz inequality that
\[
1 \leq \|K\|\|e\|
\]
and hence
\[
\|K\| \geq \frac{1}{\|e\|}. 
\]
However, if we choose 
\begin{equation}
\label{K}
K = \frac{e^T}{\|e\|^2}
\end{equation}
then (\ref{Kconstraint}) is satisfied and $\|K\| = \frac{1}{\|e\|}$. Hence, this value of $K$ must be the optimal $K$. 

We now consider the special case of $\omega_o = 0$. In this case, we obtain
\[
e =  2 \sqrt{\kappa}\left[\begin{array}{ll} \frac{2}{\kappa} & 0 \\0 & \frac{2}{\kappa} \end{array}\right] J\beta = \frac{4}{\sqrt{\kappa}}J\beta.
\]
Hence, as $\kappa \rightarrow 0$, $\|e\| \rightarrow \infty$ and therefore $\|K\| \rightarrow 0$. This means that we can make the noise level on our measurement arbitrarily small by choosing $\kappa > 0$ sufficiently small. However, as $\kappa$ gets smaller, the system (\ref{augmented4}) gets closer to instability and hence, takes longer to converge to steady state.

\section{A Possible Implementation of the Plant Observer System}
\label{sec:implement}
In this section, we describe one possible experimental implementation of the plant-observer system given in the previous section. The plant-observer system is a linear quantum system  with Hamiltonian 
\begin{equation}
\label{total_H}
\mathcal{H}_c+\mathcal{H}_o = x_{p}\trp R_{c} x_{o}+\half x_o\trp R_o x_o
\end{equation}
and coupling operator defined so that
\[
\left[\begin{array}{c} L+ L^* \\ \frac{ L - L^*}{i}\end{array}\right]
 = W_o x_o.
\]
Furthermore, we assume that $R_c = \alpha\beta^T$, $R_o = \omega_o I$, $W_o = \sqrt{\kappa} I$ where $\alpha  \in \rbb^{2}$, $\beta \in \rbb^{2}$,  $\omega_o > 0$ and $\kappa > 0$. 

\begin{figure}[htbp]
\begin{center}
\includegraphics[width=8cm]{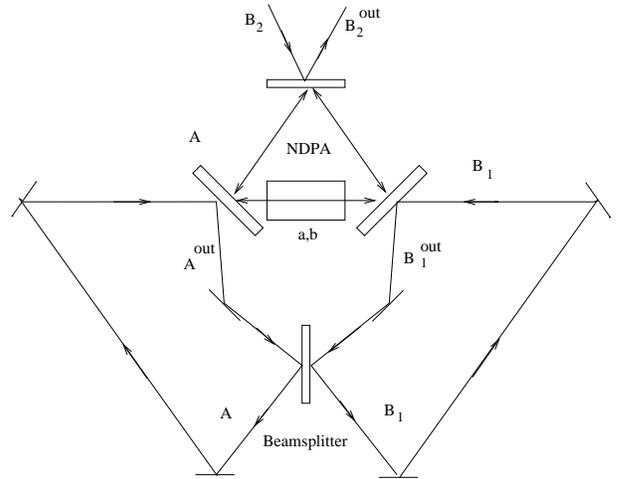}
\end{center}
\caption{NDPA coupled to a beamsplitter.}
\label{F2}
\end{figure}

In order to construct a linear quantum system with a Hamiltonian of this form, we consider an NDPA coupled to a beamsplitter as shown schematically in Figure \ref{F2}; e.g., see \cite{BR04}.

A linearized approximation for the NDPA is defined by a quadratic Hamiltonian of the form
\[
\mathcal{H}_1 = \frac{\imath}{2} \left(\epsilon a^*b^* - \epsilon^* a b\right)+\half\omega_o b^*b
\]
where $a$ is the annihilation operator corresponding to the first mode of the NDPA and $b$ is the annihilation operator corresponding to the second mode of the NDPA. These modes will be assumed to be of the same frequency but with a different polarization with $a$ corresponding to the quantum plant and $b$ corresponding to the quantum observer. Also, $\epsilon$ is a complex parameter defining the level of squeezing in the NDPA and $\omega_o$ corresponds to the detuning frequency of the $b$ mode in the NDPA. The $a$ mode in the NDPA is assumed to be tuned. In addition, the NDPA corresponds to a vector of coupling operators $L = \left[\begin{array}{l} \sqrt{\kappa_1} a \\ \sqrt{\kappa_2}b \\ \sqrt{\kappa_3}b\end{array} \right]$. Here $\kappa_1 > 0$, $\kappa_2 > 0$, $\kappa_3 >0$ are scalar parameters determined by the reflectance of the mirrors in the NDPA. 

From the above Hamiltonian and coupling operators, we can calculate the following quantum stochastic differential equations (QSDEs) describing the NDPA:
\begin{eqnarray}
\label{qsde}
\left[\begin{array}{l} d a \\ d b \end{array} \right] &=& 
 \left[\begin{array}{ll} 0 & \frac{\epsilon}{2}  \\ \frac{\epsilon}{2} & 0 \end{array} \right]\left[\begin{array}{l} a^* \\ b^* \end{array} \right]dt \nonumber \\
&&-
 \left[\begin{array}{ll} \frac{\gamma_1}{2} &  0 \\ 0 & \frac{\gamma_2}{2}+\half\imath \omega_o \end{array} \right]\left[\begin{array}{l} a \\ b \end{array} \right]dt \nonumber \\
&&- \left[\begin{array}{lll} \sqrt{\kappa_1} &  0 & 0\\ 0 & \sqrt{\kappa_2} & \sqrt{\kappa_3} \end{array} \right]\left[\begin{array}{l} dA \\ dB_1 \\ dB_2 \end{array} \right];\nonumber \\
 \left[\begin{array}{l} dA^{out} \\ dB_1^{out}\\ dB_2^{out} \end{array} \right] &=& \left[\begin{array}{ll} \sqrt{\kappa_1} &  0 
\\ 0 & \sqrt{\kappa_2} \\ 0 & \sqrt{\kappa_3}\end{array} \right]
\left[\begin{array}{l} a \\ b \end{array} \right]dt + \left[\begin{array}{l} dA \\ dB_1\\ dB_2 \end{array} \right];\nonumber \\
\end{eqnarray}
where $\gamma_1 = \kappa_1$ and $\gamma_2 = \kappa_2+\kappa_3$.

 We now consider the equations defining the beamsplitter
 \[
 \left[\begin{array}{l} A \\ B_1 \end{array} \right] = \left[\begin{array}{ll} \cos \theta  & e^{-\imath \phi} \sin \theta  \\ -e^{\imath \phi} \sin \theta  & \cos \theta \end{array} \right]\left[\begin{array}{l} A^{out} \\ B_1^{out} \end{array} \right]
 \]
 where $\theta$ and $\phi$ are angle parameters defining the beamsplitter; e.g., see \cite{MW95}. This implies
 \[
 \left[\begin{array}{l} A^{out} \\ B_1^{out} \end{array} \right] = \left[\begin{array}{ll} \cos \theta  & -e^{-\imath \phi} \sin \theta  \\ e^{\imath \phi} \sin \theta  & \cos \theta \end{array} \right] \left[\begin{array}{l} A \\ B_1 \end{array} \right].
 \]
 Substituting this into the second equation in (\ref{qsde}), we obtain
 \begin{eqnarray*}
 \lefteqn{\left[\begin{array}{ll} \cos \theta  & -e^{-\imath \phi} \sin \theta  \\ e^{\imath \phi} \sin \theta  & \cos \theta \end{array} \right] \left[\begin{array}{l} dA \\ dB_1 \end{array} \right] }\nonumber \\
&&= \left[\begin{array}{ll} \sqrt{\kappa_1} &  0\\ 0 & \sqrt{\kappa_2}  \end{array} \right] \left[\begin{array}{l} a \\ b \end{array} \right]dt + \left[\begin{array}{l} dA \\ dB_1 \end{array} \right]
 \end{eqnarray*}
and hence
\begin{eqnarray*}
\left[\begin{array}{ll} \cos \theta -1   & -e^{-\imath \phi} \sin \theta  \\ e^{\imath \phi} \sin \theta  & \cos  \theta -1 \end{array} \right] \left[\begin{array}{l} dA \\ dB_1 \end{array} \right]\\
=\left[\begin{array}{ll} \sqrt{\kappa_1} &  0\\ 0 & \sqrt{\kappa_2}  \end{array} \right] \left[\begin{array}{l} a \\ b \end{array} \right]dt.
\end{eqnarray*}
We now assume that $\cos \theta \neq 1$. It follows that we can write
\begin{eqnarray*}
\lefteqn{\left[\begin{array}{l} dA \\ dB_1 \end{array} \right] =}\nonumber \\
&& \frac{1}{2(1-\cos \theta)}\left[\begin{array}{ll} \cos \theta -1   & e^{-\imath \phi} \sin \theta  \\ -e^{\imath \phi} \sin \theta  & \cos  \theta -1 \end{array} \right]\nonumber \\
&&\times \left[\begin{array}{ll} \sqrt{\kappa_1} &  0\\ 0 & \sqrt{\kappa_2}  \end{array} \right]\left[\begin{array}{l} a \\ b \end{array} \right]dt.
\end{eqnarray*}
Substituting this into the first equation in (\ref{qsde}), we obtain
\begin{eqnarray*}
&&\left[\begin{array}{l} d a \\ d b \end{array} \right]=\nonumber \\
 &&\left[\begin{array}{ll} 0 & \frac{\epsilon}{2}  \\ \frac{\epsilon}{2} & 0 \end{array} \right]\left[\begin{array}{l} a^* \\ b^* \end{array} \right]dt \nonumber \\
&&-
 \left[\begin{array}{ll} \frac{\gamma_1}{2} &  0 \\ 0 & \frac{\gamma_2}{2}+\half\imath \omega_o \end{array} \right]\left[\begin{array}{l} a \\ b \end{array} \right]dt \nonumber \\
&&- \frac{1}{2(1-\cos \theta)}\left[\begin{array}{ll} 
\sqrt{\kappa_1} &  0\\ 0 & \sqrt{\kappa_2}  \end{array} \right]
\left[\begin{array}{ll} \cos \theta -1   & e^{-\imath \phi} \sin \theta  \\ -e^{\imath \phi} \sin \theta  & \cos  \theta -1 \end{array} \right]\nonumber \\
&& \times \left[\begin{array}{ll} \sqrt{\kappa_1} &  0\\ 0 & \sqrt{\kappa_2}  \end{array} \right]
\left[\begin{array}{l} a \\ b \end{array} \right]dt \nonumber \\
&&-\left[\begin{array}{l}  0\\  \sqrt{\kappa_3} \end{array} \right]dB_2; \nonumber \\
&&dB_2^{out} = \sqrt{\kappa_3}b + dB_2.
\end{eqnarray*}
These QSDEs can be written in the form
\begin{eqnarray*}
\left[\begin{array}{l} d a \\ d b \\da^* \\ db^*\end{array} \right] &=& F \left[\begin{array}{l} a \\ b \\a^* \\ b^*\end{array} \right]dt 
+ G \left[\begin{array}{l}dB_2\\dB_2^*\end{array} \right]; \nonumber \\
\left[\begin{array}{l}dB_2^{out}\\dB_2^{out*}\end{array} \right] &=& H\left[\begin{array}{l} a \\ b \\a^* \\ b^*\end{array} \right]dt  
+ \left[\begin{array}{l}dB_2\\dB_2^*\end{array} \right]
\end{eqnarray*}
where the matrix $F$ is given by $F =\left[\begin{array}{ll}F_1 & F_2 \\F_2^\# & F_1^\#\end{array}\right]$ and
\begin{eqnarray*}
F_1 &=& \left[\begin{array}{ll}0 & - \frac{\sqrt{\kappa_1\kappa_2} e^{-\imath \phi} \sin \theta}{2(1-\cos \theta)}\\
\frac{\sqrt{\kappa_1\kappa_2} e^{\imath \phi} \sin \theta}{2(1-\cos \theta)} & -\frac{\kappa_3}{2}-\half \imath \omega_o\end{array}\right],\\
F_2 &=& \left[\begin{array}{ll}0 & \frac{\epsilon}{2}\\\frac{\epsilon}{2} & 0\end{array}\right].
\end{eqnarray*}
Also, the matrix $G$ is given by
\[
G=-\left[\begin{array}{ll}  0 & 0\\  \sqrt{\kappa_3} & 0 \\ 0 & 0\\ 0 &  \sqrt{\kappa_3}\end{array} \right],
\]
and the matrix $H$ is given by
\[
H= \left[\begin{array}{llll}0 & \sqrt{\kappa_3} & 0 & 0\\0 & 0 & 0 &\sqrt{\kappa_3}\end{array} \right].
\]
It now follows from the proof of Theorem 1 in \cite{ShP5} that we can construct a Hamiltonian for this system of the form
\[
\mathcal{H} = \half \left[\begin{array}{llll} a^*& b^*&a & b\end{array} \right]M\left[\begin{array}{l} a \\ b \\a^* \\ b^*\end{array} \right]
\]
where the matrix $M$ is given by 
\[
M = \frac{\imath}{2}\left(JF-F^\dagger J\right)
\]
and $J= \left[\begin{array}{ll}I & 0 \\0 & -I\end{array} \right]$. Then, we calculate $M =\left[\begin{array}{ll}M_1 & M_2 \\M_2^\# & M_1^\#\end{array}\right]$ where
\begin{eqnarray*}
M_1 &=& \frac{\imath}{2}\left[\begin{array}{ll}0 & - \frac{\sqrt{\kappa_1\kappa_2} e^{-\imath \phi} \sin \theta}{1-\cos \theta}\\
\frac{\sqrt{\kappa_1\kappa_2} e^{\imath \phi} \sin \theta}{1-\cos \theta} & -\imath \omega_o\end{array}\right],\nonumber \\
M_2 &=& \frac{\imath}{2}\left[\begin{array}{ll}0 & \epsilon\\\epsilon & 0\end{array}\right].
\end{eqnarray*}

Also, we can construct the coupling operator for this system in the form
\[
L = \left[\begin{array}{ll}N_1 & N_2 \end{array}\right]\left[\begin{array}{l} a \\ b \\a^* \\ b^*\end{array} \right]
\]
where the matrix $N=\left[\begin{array}{ll}N_1 & N_2\\N_2^\# & N_1^\# \end{array}\right]$ is given by 
\[
N=H.
\]
Hence,
\[
N_1 = \left[\begin{array}{ll}0 & \sqrt{\kappa_3} \end{array} \right], N_2 = 0. 
\]
We now wish to calculate the Hamiltonian $\mathcal{H}$ in terms of the quadrature variables defined such that 
\[
\left[\begin{array}{l} a \\ b \\a^* \\ b^*\end{array} \right] = 
\Phi \left[\begin{array}{l} q_p \\ p_p \\q_o \\ p_o\end{array} \right]
\]
where the matrix $\Phi$ is given by
\[
\Phi = \left[\begin{array}{llll} 
1 & \imath & 0 & 0 \\
0 & 0 & 1 & \imath \\
1 & -\imath & 0 & 0 \\
0 & 0 & 1 & -\imath
\end{array} \right].
\]
Then we calculate
\begin{eqnarray*}
\mathcal{H} &=& \half \left[\begin{array}{llll}q_p & p_p &q_o & p_o \end{array} \right]R\left[\begin{array}{l} q_p \\ p_p \\q_o \\ p_o\end{array} \right]\nonumber \\
&=& \half \left[\begin{array}{ll}x_p\trp & x_o\trp \end{array} \right] R\left[\begin{array}{l}x_p \\ x_o \end{array} \right]
\end{eqnarray*}
where the matrix $R$ is given by
\begin{eqnarray*}
R &=& \Phi^\dagger M \Phi\\
&=& \left[\begin{array}{ll} 0 & R_c \\
R_c\trp & \omega_o I
\end{array} \right],
\end{eqnarray*}
\[
R_c = \left[\begin{array}{ll}
-\Im(\epsilon) -\Im(\delta) & \Re(\epsilon)+\Re(\delta) \\
\Re(\epsilon) -\Re(\delta) & \Im(\epsilon)-\Im(\delta)
\end{array} \right]
\]
and $\delta = \frac{\sqrt{\kappa_1\kappa_2} e^{\imath \phi} \sin \theta}{1-\cos \theta}$. Hence,
\[
\mathcal{H} = \half\omega_o x_{o}\trp x_{o}+ x_{p}\trp R_{c} x_{o}.
\]
Comparing this with equation (\ref{Rc}), we require that 
\begin{equation}
\label{Rc1}
\left[\begin{array}{ll}
-\Im(\epsilon) -\Im(\delta) & \Re(\epsilon)+\Re(\delta) \\
\Re(\epsilon) -\Re(\delta) & \Im(\epsilon)-\Im(\delta)
\end{array} \right] = \alpha\beta^T
\end{equation}
and the condition (\ref{Kconstraint})  to be satisfied in order for the system shown in Figure \ref{F2} to provide an implementation of the augmented plant-observer system. 

\begin{figure}[htbp]
\begin{center}
\includegraphics[width=8cm]{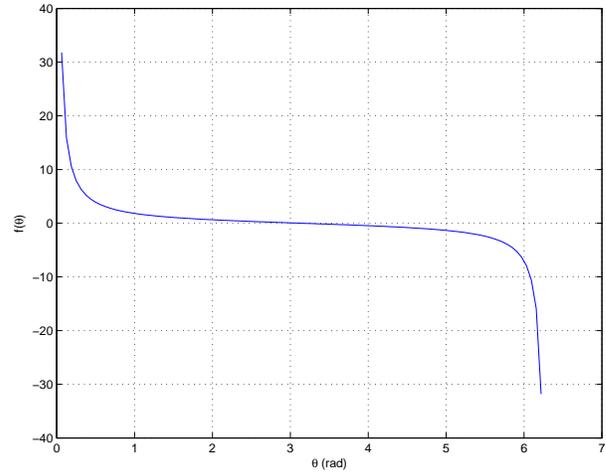}
\end{center}
\caption{Plot of the function $f(\theta)$.}
\label{F4}
\end{figure}

We first observe that the matrix on the right hand side of equation (\ref{Rc1}) is a rank one matrix and hence, we require that
\[
\det \left[\begin{array}{ll}
-\Im(\epsilon) -\Im(\delta) & \Re(\epsilon)+\Re(\delta) \\
\Re(\epsilon) -\Re(\delta) & \Im(\epsilon)-\Im(\delta)
\end{array} \right] = |\delta|^2 - |\epsilon|^2 = 0.
\]
That is, we require that 
\[
\sqrt{\kappa_1\kappa_2} \left|\frac{\sin \theta}{1-\cos \theta}\right| = |\epsilon|. 
\]
Note that the function $\frac{\sin \theta}{1-\cos \theta}$ takes on all values in $(-\infty,\infty)$ for $\theta \in (0,2\pi)$ and hence, this condition can always be satisfied for a suitable choice of $\theta$. This can be seen in Figure \ref{F4} which shows a plot of the function $f(\theta) = \frac{\sin \theta}{1-\cos \theta}$.

Furthermore, we will assume without loss of generality that $\theta \in (0,\pi)$ and hence we obtain our first design equation
\begin{equation}
\label{theta_eqn}
\frac{\sin \theta}{1-\cos \theta} = \frac{|\epsilon|}{\sqrt{\kappa_1\kappa_2}}.
\end{equation}
In practice, this ratio would be chosen in the range of $\frac{|\epsilon|}{\sqrt{\kappa_1\kappa_2}} \in (0,0.6)$ in order to ensure that the linearized model which is being used is valid. 

We now construct the vectors $\alpha$ and $\beta$ so that condition (\ref{Rc1}) is satisfied. Indeed, we let
\[
\alpha = \left[\begin{array}{l}1\\\frac{\Re(\epsilon)-\Re(\delta)}{-\Im(\epsilon) -\Im(\delta)}\end{array}\right],~~
\beta= \left[\begin{array}{l} -\Im(\epsilon) -\Im(\delta)\\\Re(\epsilon) +\Re(\delta)\end{array}\right].
\]
For these values of $\alpha$ and $\beta$, it is straightforward to verify that (\ref{Rc1}) is satisfied provided that $|\epsilon| = |\delta|$.
With this value of $\beta$, we now calculate the quantity $e$ defined in (\ref{e}) as follows:
\begin{eqnarray*}
e&=&\frac{4}{\kappa_3^2+16 \omega_o^2}\left[\begin{array}{ll} \kappa_3 & 4 \omega_o \\-4\omega_o &  \kappa_3 \end{array}\right]J\beta \\
&=& -\frac{4}{\kappa_3^2+16 \omega_o^2}\left[\begin{array}{ll} \kappa_3 & 4 \omega_o \\-4\omega_o &  \kappa_3 \end{array}\right]
\left[\begin{array}{l}\Re(\epsilon) +\Re(\delta) \\\Im(\epsilon)+ \Im(\delta)\end{array}\right].
\end{eqnarray*}
Then, the vector $K$ defining the quadrature measured by the homodyne detector is constructed according to the equation (\ref{K}).

In the special case that $\omega_o =0$, this reduces to 
\[
e= -\frac{4}{\kappa_3}\left[\begin{array}{l}\Re(\epsilon) +\Re(\delta) \\\Im(\epsilon)+ \Im(\delta)\end{array}\right].
\]
In terms of complex numbers ${\bf e} = e(1)+\imath e(2)$, we can write this as
\[
{\bf e} = -\frac{4}{\kappa_3}(\epsilon + \delta). 
\]
Then, in terms of complex numbers ${\bf K} = K(1)+\imath K(2)$, the formula (\ref{K}) becomes
\[
{\bf K} = \frac{{\bf e}}{|{\bf e}|^2} = -\frac{\kappa_3(\epsilon+\delta)}{4(\epsilon+\delta)(\bar \epsilon+\bar \delta)}
 = -\frac{\kappa_3}{4(\bar \epsilon+\bar \delta)}
\]
where $\bar{(\cdot)}$ denotes complex conjugate. Also, as noted in Section \ref{sec:observer}, the steady state measurement noise intensity is given by
\[
\frac{1}{\|e\|} = -\frac{\kappa_3}{4|\epsilon+ \delta|}
\]
which approaches zero as $\kappa_3 \rightarrow 0$. However, this is at the expense of increasingly slower convergence to steady state. 


\section{Conclusions}
In this paper, we have shown that a direct coupling observer for a linear quantum system can be implemented in the case that the observer can be measured using a Homodyne detection measurement. This would allow the plant observer system to be constructed experimentally and the performance of the observer could be verified using the measured data.


\end{document}